\documentclass[%
reprint,
amsmath,amssymb,
aps,
prb,
titleinbib,
]{revtex4-2}

\usepackage{graphicx}
\usepackage{dcolumn}
\usepackage{bm}
\usepackage{braket}
\usepackage{float}
\usepackage{cancel}
\usepackage[dvipsnames]{xcolor}
\newcommand{\corr}{\textcolor{black}}
\newcommand{\improv}{\textcolor{black}}
\newcommand{\last}{\textcolor{black}}

\begin{document}

\preprint{APS/123-QED}

\title{Tailoring transport in quantum spin chains via disorder and collisions}

\author{Vittoria Stanzione}
\author{Alessandro Civolani}%
\altaffiliation{ Physics Department and SUPA, Strathclyde University, Glasgow G4 0NQ, UK }%
\author{Jorge Yago Malo}
 \email{corresponding author: jorge.yago@unipi.it}
\author{Maria Luisa Chiofalo}
\affiliation{Physics Department and INFN, Pisa University, Pisa, I-56127, Italy .}

\date{\today}

\begin{abstract}
 We systematically investigate the interplay of disorder and time-homogeneous collisional noise in shaping the transport dynamics of an anisotropic XXZ spin chain. Using stochastic collision models to simulate interaction with the environment, we explore the localization-delocalization transitions across regimes with single and multiple excitations. We find that space-homogeneous and low-rate collisions can shape regions where localization sets in the form of subsequent plateaus. The localization process has universal features for the plateaus duration and the delocalization time. Interactions among the excitations favor this process even for tiniest disorder levels.   

Our findings can be applied to design stroboscopic protocols where sequences of transport and localization can be tailored. We establish relevant connections to noise-engineering of quantum devices in noisy intermediate-scale quantum simulators platforms, and to realistic biological systems where noise and disorder coexist. 

\end{abstract}

\maketitle


\section{\label{sec:intro}Introduction\\}

Quantum transport governs charge, spin and energy transfer in a wide variety of systems, from solid-state physics and spintronics to quantum technologies \cite{YagoMalo2024, Ferry_2022, Brahlek2024, Sebolt2024}. In fact,  by exploring the effects of disorder, noise, and interactions, we can gain insights into the mechanisms driving localization-delocalization transitions, thermalization, and the emergence of collective phenomena in low-dimensional and complex systems. In particular, current technologies enable the study of quantum transport with the engineering of reduced-dimensionality systems in solid state and in AMO (atomic, molecular, and optical) platforms \cite{Wili2023, Huang2024}.

We are interested in transport of energy and excitations in biological systems~\cite{Plenio2008, SCM_approach_to_transport} and in coarse-grained models within the so-called quantum-like paradigm \cite{YagoMalo2023, Khrennikov}. 
Regardless of the presence of quantum effects in biological processes, the open quantum systems framework \cite{non_markov_review, Wiseman_Milburn, Petruccione_Breuer, Gardiner_Zoller} offers a well-suited mathematical toolbox for modeling complex phenomena, both quantum and classical \cite{AlessandroetAl, Khrennikov}. Interestingly, open quantum networks modeled by paradigmatic many-body Hamiltonians can also be engineered in tunable current quantum simulators to reproduce essential transport dynamics \cite{roy2024simulating}.

Quantum effects in biological systems present a fascinating yet debated area of research, challenging conventional views on the role of quantum mechanics in the warm and noisy environment of living organisms. These physics-based models can incorporate quantum effects into the description of microscopic phenomena, allowing us to study how a noisy and out-of-equilibrium biological environment is influenced by coherent and dissipative quantum effects  \cite{Plenio2008, OlayaCastro2008,pathania2024dynamics}. 

In general, the behavior of transport in a quantum system serves as a powerful diagnostic tool for understanding thermalization in disordered conditions \cite{dohner2023thermalization}. In the absence of interactions, disorder can induce Anderson localization \cite{Guan2019}, where transport is suppressed due to interference effects, preventing the system from reaching thermal equilibrium. Interactions or noise disrupts this localization mechanism, leading to a transition toward diffusive or even ballistic transport, depending on the system's parameters \cite{MBL_Nandkishore,MBL_from_IPR_study}. The transition from localized to delocalized regimes provides insight into how thermalization emerges in complex systems. Consequently, disorder is a mechanism that suppresses transport in isolated systems and, in open systems, also a factor that competes with noise to dictate the system's relaxation dynamics.

In \cite{AlessandroetAl}, we explored how dissipation modulates transport in an open quantum system using stochastic collision models accounting for space and time heterogeneous and homogeneous collisions. In so doing, we could focus on the interplay between noise, interactions, and coherence under sufficiently general conditions. By tuning noise parameters, we demonstrated controllable modifications to transport properties, identifying regimes where increasing the collision rate enhances transport. Thus, we  showed that engineered dissipation can counteract localization effects.

We now focus on the combination of collisional noise and disorder which provides a model that closely mirrors the dynamics of transport in many complex media including those relevant to biological systems. Collisional models replicate approximate interactions with noisy warm environment, while disorder introduces local spatial inhomogeneities. Together, these elements capture key features of energy transfer in systems where coherent and dissipative processes coexist, relevant in several fields of physics. Examples are charging and discharging performances in open quantum system batteries \cite{Carrega2020, Mayo2022}, transport in an engineered environment of complex systems \cite{Nokkala2024} along with many body localization in 1-D systems \cite{Levi2016}.  Besides, the tunability of collisional noise parameters enables the simulation of biologically relevant scenarios, such as the exploitation of noise-assisted transport \cite{Huelga2009, Huelga2013}.

In this paper, we investigate the transport properties and thermalization dynamics of a noisy, disordered XXZ spin chain \cite{Prosen_Buca_correlations_XXZ,Prosen_Buca_XXZ, Rakov_Weyrauch_Braiorr-Orrs, McRoberts2024,KT_XXZ}, modeled through stochastic collision processes. Among different possibilities,  we adopt this many-body model for its capability of describing in a general manner tunneling-based transport of excitations in the presence of coherent anisotropic interactions. We identify localization/delocalization transitions and thus transport windows for multiple excitations tuning the interplay of disorder, anisotropy, and noise. 
 
 The paper is organized as follows. In Section \ref{sec:model} we present the model of our disordered spin-chain interacting with the environment via  stochastic collision models and introduce the figures of merit adopted to study transport properties. In Section~\ref{Res}, we present the results of our study comparing the cases of one, two and multiple injected excitations also highlighting the entanglement properties in those regimes. Finally, in Section~\ref{Discc} we present our conclusions and discuss future perspectives in the light of technological progress in quantum science and biology.

 \section{\label{sec:model}Model and methods\\}

\begin{figure}[htbp]
\includegraphics[scale = 0.16]{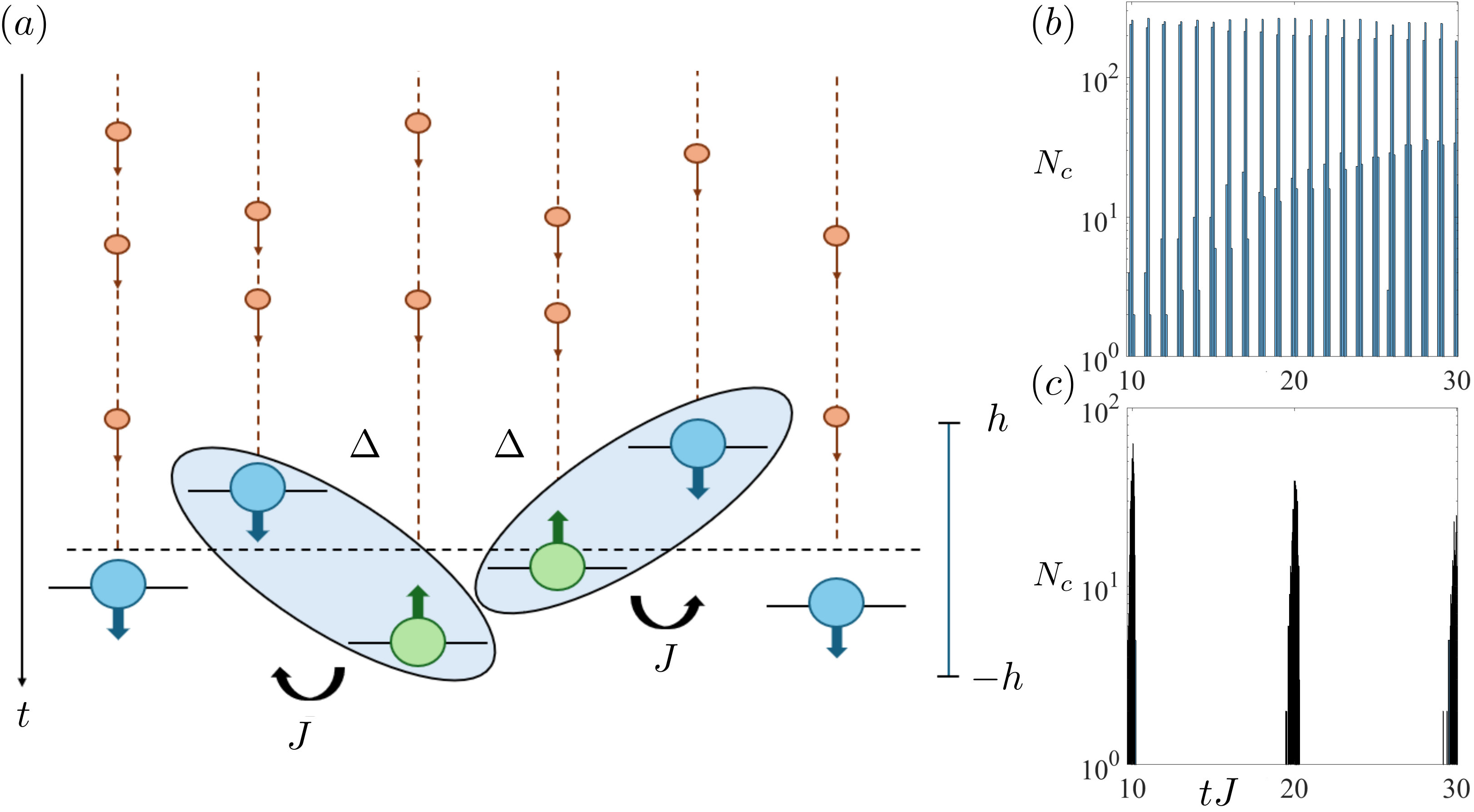}
\caption{ Sketch of our model. (a) Open quantum spin chain with energy levels affected by disorder in the $[-h, h]$ range. Excitations are depicted as flipped spins (green). Their dynamics is affected by the tunneling rate $J$ (black arrows), interaction strength $\Delta$ (blue ovals) and the addition of a noisy environment. The latter is represented by the auxiliary light-brown qubits colliding with the sites over time. (b)-(c) Histograms displaying the number of collisions undergone over time tJ by the auxiliary qubits (log scale) against spins. This is shown for a noise shape parameter $\nu = 100$ in the time-uniform regime, and collision rates $r_c = 1$ (b) and (c) $r_c = 0.1$. We use the same simulation parameters: number of sites in the spin is $N = 41$ sites, $M=500$ trajectories and final time of simulation $tJ = 30$.}
    \label{fig:Model}
\end{figure}

The conceptual model setting is sketched in Fig.~\ref{fig:Model} (a). We investigate an integrable generalization of the Heisenberg spin chain that accounts for uni-axial anisotropy \cite{Franchini_chains}, with the addition of a disorder term:
\begin{eqnarray}
    H_{XXZ}=J \sum_{i=1}^{N-1} \left[ \sigma^x_i \sigma^x_{i+1} + 
    \sigma^y_i \sigma^y_{i+1} + 
    \Delta \sigma^z_i \sigma^z_{i+1} \right] \nonumber \\ 
    +  \sum_{i=1}^{N} h_i\sigma^z_{i},
    \label{XXZ_Hamiltonian}
\end{eqnarray}
and subjected to dissipation through noise in the form of stochastic collision models, as depicted in Sec.~\ref{sec:Noise_des}, with open boundary conditions (OBC). In Eq.~\eqref{XXZ_Hamiltonian}, $\sigma^{x/y/z}_{i}$ are the Pauli operators acting on spin located at site $i$; $J$ is the spin exchange rate, that can also be seen as the tunneling amplitude for excitation quasiparticles throughout the system; $\Delta$ governs the anisotropy in the form of a spin-spin interaction along the z-axis. The last term introduces disorder for every spin independently. This is in the form of a random local field, where each $h_i$ is drawn individually from the uniform distribution over the interval
$[-h, h]$. In our analysis, we consider the regime with $J=1 (>0)$ and then scale every other energy to $J$. As a result, the ground state with disorder $h=0$ is an anti-ferromagnet for $\Delta>1$, and a paramagnet for $0 < \Delta < 1$.
When considering the time evolution of excitations or quench dynamics, the disorder term will control different transport regimes, ranging from ballistic to diffusive, and it can lead to the emergence of Many-Body localization (MBL) \cite{Prosen2011}. In fact, disorder can lead to the breakdown of thermalization, meaning that the system fails to reach thermal equilibrium at long times resulting in transport suppression. Tuning the disorder through $h$ and analyzing the resulting transport behavior, it is possible to map the parameter regimes where localization or delocalization dominates.

\subsection{\label{sec:Noise_des}Noise modeling}

So far we have discussed how the coherent and disorder-driven terms of our model are described. Here, we introduce dissipation through a particular discretization of the environment degrees of freedom (d.o.f.) that lies in the family of stochastic collision models \cite{SCM_approach_to_transport,Gallina}.
This type of environment description captures spatio-temporal homogeneity and heterogeneity in the system-bath interaction, effectively approximating biological environments \cite{SCM_approach_to_transport, Cao2020}. As depicted in Fig.~\ref{fig:Model}, the environment d.o.f are modeled as auxiliary qubits (light brown circles) that collide over time with the spins in the chain.  

In this approach, we can govern the noise regimes via appropriate distributions describing the probability of system-environment interaction events. This results in a flexible environment description, including non-trivial temporal correlations even remaining within a Markovian-like description. As it is the case for many dissipative dynamical maps, stochastic collisional models are compatible with many local quantum channels. In our case, we account for density-density collisions decorrelating spins from the coherent evolution, thus effectively producing dephasing in the system degree of freedom.

We choose to describe the collisional rate between spins and auxiliary d.o.f. by adopting the especially flexible Weibull distribution \cite{Weibull_Renewal_Processes,SCM_approach_to_transport}:
\begin{equation}
    p(t)=\frac{\nu}{\mu}
    \left(
    \frac{t}{\mu}
    \right)^{\nu-1}
    e^{-(t/\mu)^\nu},
    \label{Weibull_noise}
\end{equation}
where indeed the shape parameter $\nu \geq 0$ controls the time heterogeneity structure,  and the scale parameter $\mu > 0$ governs its overall rate. Specifically, collisions are heterogeneous over time for $\nu \leq 1$ and homogeneous for $\nu \gg 1$ while the intercollision time becomes constant.
 In contrast, the scale parameter is related to the overall collision rate that we define below:
\begin{equation}
    r_c=\frac{1}{\tau_{th}}=\frac{1}{\mu\:\Gamma\:(1+1/\nu)},
\end{equation}
with $\tau_{th}$ being the mean collision time and $\Gamma$ given by $\int_0^\infty \tau p(\tau) d\tau= \mu\Gamma(1+1/\nu)=t_{th}$ from Eq.~\eqref{Weibull_noise}). 
In Fig.~\ref{fig:Model}. (b)-(c), we present histograms for the number of collisions as a function of time for homogeneous noise ($\nu = 100$) and two different collision rates modifying how recurrent is the interaction with the environment. Regarding the description of noise, we specify here that we use the same nomenclature as \cite{SCM_approach_to_transport} where the temporal heterogeneity is defined by looking at the variance
of the stochastic process and the regularity with which the collisions
occur.

\corr{The collision times are generated independently for each site as follows: given a number taken randomly from the uniform distribution $u \in [0,1]$, by inverse-transforming sampling, the waiting time since the previous collision is going to be $ t = \sqrt[\nu]{-\mu^{\nu} \ln\!\left|1-u\right|}$,
with $\nu$ and $\mu$ being the shape and scale parameter of Eq.~\ref{Weibull_noise}. 
The initial list of collision times $\bar{S}$ contains these first waiting times (interpreted as absolute times of the first collisions) for each site. After a collision occurs at site $i$ at time $S_i$, a new independent waiting time is drawn in the same way and added to $S_i$, so that $S_i \to S_i+t_{\text{new}}$. Thus, the next site in which a collision will occur will be simply given by the minimum of the updated list of waiting times $\bar{S}'$.}

When collisions occur according to the probability distribution, a quantum channel is applied~\cite{SCM_approach_to_transport,Decoherence_without_entanglement_Quantum_Darwinism}. This is given by: 

 \begin{equation}\label{eq:quantum_channel}
     \Phi\left[\rho(t_i)\right]=\textrm{Tr}[ U_{\textrm{coll}}(\rho_a\otimes\rho(t_i))U^\dagger_{\textrm{coll}}]\,,
 \end{equation}
 with $U_{\textrm{coll}}=\exp{-i(\pi/2)\sigma_a^x\otimes\sigma^z_i}$ representing the collision event and $t_i$ being the time before the given collision. Initially, the system is prepared in a well-defined state, which typically consists of one or more spins flipped along the chain. Subsequently, for each site of the system, a list of future collision times is generated according to a stochastic process governed by the chosen noise distribution.
Between two consecutive collisions, the system evolves coherently, according to the dynamics dictated by the Hamiltonian in Eq.\eqref{XXZ_Hamiltonian}. When a collision occurs at a given site, the system undergoes a transformation described by the quantum channel in Eq.\eqref{eq:quantum_channel}.  

After the application of the quantum channel, the time for the next collision at the affected site is resampled from the same probability distribution. The entire procedure is then iteratively repeated, allowing for the simulation of the time evolution of the system\corr{, using exact diagonalization,} in the presence of structured noise. For a detailed description, both of the noise model and
the effects of the collisions, we refer to our previous work \cite{AlessandroetAl}.
\corr{We note here that we kept the same disorder realization across different values of the parameters, to isolate the consequences of changes due to collisions. We have verified that our conclusions remain unchanged under different disorder realizations.}

Consequently, three competing mechanisms act on the spin chain: coherent coupling between spins, local disorder, and noise originating from the Weibull distribution.

\subsection{\label{FOM}Characterizing transport}

Having described our model, we now introduce the figures of merit that we use to characterize the system transport and quantify the magnetization spreading (delocalization behavior of the excitations). For one excitation, we use the inverse participation ratio (IPR), $\mathrm{IPR}= \sum_{i=1}^N\bra{i}\rho(t)\ket{i}^2, $
where $i$ represents the site index and $\ket{i}$  represents the single-excitation localized states $  \Big\{
    \ket{i}\: : \:\ket{i}=\sigma_{i}^{\dagger} \ket{0}^{\otimes N}
    \Big\} $ \cite{SCM_approach_to_transport}. The IPR  is a measure of localization, bounded between the complete delocalization asymptotic value $\mathrm{IPR}=1/N$, and $\mathrm{IPR}=1$, corresponding to complete localization, i.e. when the excitation remains on a particular site of the network \cite{CTQW_spatially_correlated_noisy}. The larger the $\mathrm{IPR}$, the more localized the excitation is over the lattice. 
 
Instead, for a generic number of excitations we use the \textit{Inverse Ergodicity Ratio} (IER) $\text{IER(t)}= \sum_{j=1}^{dimH} \bra{j} \rho(t) \ket{j}, $
defined in terms of the multiple-excitation states $\ket{j}$ that compose our computational basis \cite{AlessandroetAl} $  \Big\{
    \ket{j}\: : \:\ket{j}=\sigma_{i1}^{\dagger}\sigma_{i2}^{\dagger}...\sigma_{iq-1}^{\dagger}\sigma_{iq}^{\dagger} \ket{0}^{\otimes N}
    \Big\} $, where $i_l$
 indicates the sites in which we have the q excitations. IER has the following asymptotic behaviors: $\mathrm{IER}=1/\text{dim} \mathcal{H}$ implies that the system is in a superposition of all the possible states in the reduced magnetization sector basis, thus the state is \textit{ergodic} \cite{ET_hypothesis}; $\mathrm{IER}=1$, instead, refers to the case in which the system is in one out of these specific configurations, that forms part of our basis. For more details on these two figures of merit in a similar context, we refer to \cite{AlessandroetAl}.

Our system is a disordered spin-chain, thus we expect a competition between disorder ($h$), coherent tunneling ($J$), anisotropy ($\Delta$), and noise ($r_c,\nu$) that could provide a rich landscape where localization and delocalization transitions can be explored.

Therefore, to characterize the emergence of parameter regions where the system (de)localizes, we also consider the behavior of the entanglement entropy for selected scenarios. The entanglement entropy constitutes one of the well-known heralds of MBL \cite{MBL_Nandkishore,Jorge_Localization}, as it presents logarithmic growth in many body localized systems. While not generally accessible in experiments, with a few important exceptions \cite{Inn_toolbox, Greiner}, the study of entanglement entropy in thermal and nonthermal many-body states \cite{Banuls} has helped to understand and quantify the degree of correlation and information spreading between system parts. Its behavior can for example become non-trivial in regions close to the transition point \cite{Banuls2}. The entanglement entropy can be characterized via the Von Neumann entropy \cite{Nielsen_Chuang} defined as follows: consider a bi-partite system composed by $A$ and $B$ subsystems, then the entanglement entropy is $ S_{vN} = \rho_A \log \rho_A $, with $\rho_A = Tr_B \rho$ the reduced density matrix of subsystem $A$. We are interested in regions in which transport can be modulated to occur only in well-defined time windows. Thus, we analyze the behavior of entanglement entropy as a complementary figure of merit.

\section{\label{Res}Results\\}
\subsection{\label{One_Exc}One excitation case}

We first consider the single-excitation case, so to
isolate the competition between quenched disorder (favoring
localization \cite{MBL_Nandkishore}) and noise (favoring delocalization), while avoiding
complications from interactions between excitations.

\begin{figure*} 
    \includegraphics[scale=0.27]{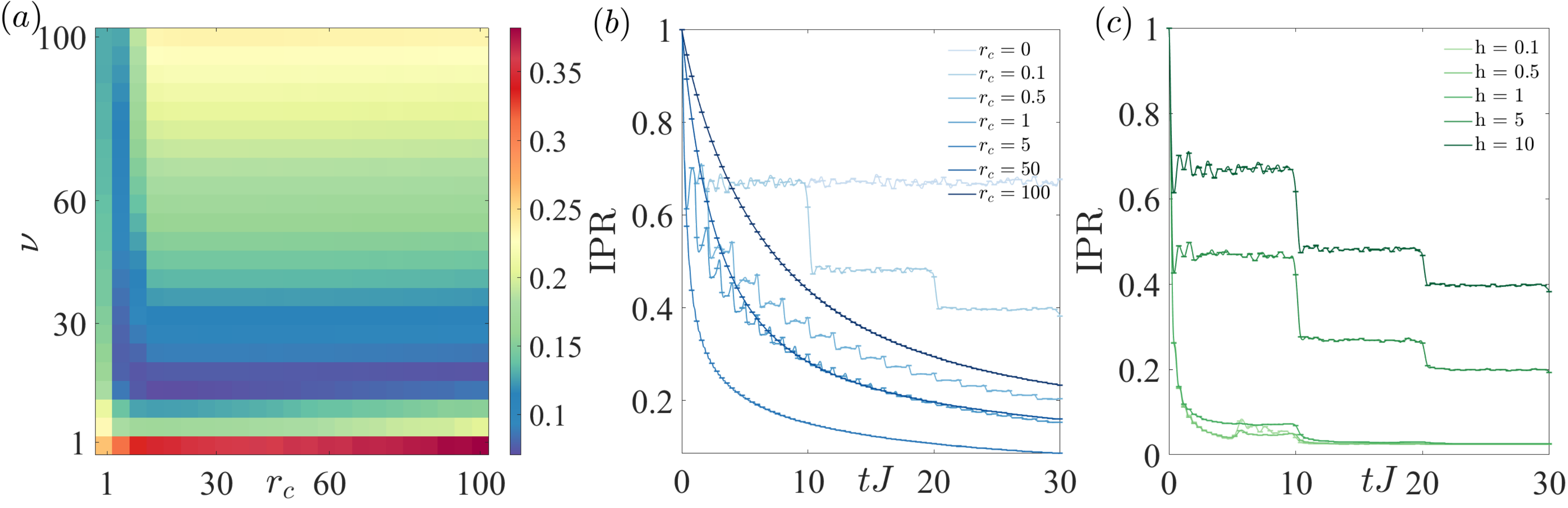}
    \caption{(De)localization behavior: the case of one excitation. Inverse Participation Ratio (IPR) for a spin chain with $N = 41$ sites. (a) IPR at long time ($tJ = 30$), for large disorder range ($h=10$) and different shape parameters $\nu$ and collision rates $r_c$. (b) IPR vs time for large disorder range ($h=10$) and different collision rates $r_c$. Curves with progressively increasing tickness and color scale refer to increasing values of $r_c=0, 0.1, 0.5, 1, 5, 50, 100$. (c) IPR vs time $tJ$ for low $r_c = 0.1$ and different disorder range $h$. Notice the emergence of plateaus of temporary stabilization of localized behavior. In (b) and (c) the shape parameter is $\nu = 100$ in the time-homogeneous collision regime. The data refer to the case of no anisotropy ($\Delta = 0$ in Eq.~\eqref{XXZ_Hamiltonian}) since they are seen to be not significantly altered by finite $\Delta \neq 0$. The simulations were performed with $M = 500$ trajectories and timestep timestep $dt = 0.02$.} 
    \label{fig:2}
\end{figure*}

We summarize in Fig.~\ref{fig:2} the IPR behavior. Fig.~\ref{fig:2}(a) summarizes the long-time behavior of the Inverse Participation Ratio (IPR) at time $tJ = 30$ as a function of the shape parameter $\nu$ and the collision rate $r_c$. We see that for collisions heterogeneous over time (low values of $\nu $) with a high rate of interaction (high $r_c$), the system exhibits higher localization (larger IPR) preventing the IPR to reach its asymptotic value in our observation window. This resembles the case of no disorder \cite{AlessandroetAl}, even if the localization is stronger here even for intermediate values of $r_c$. Instead when $\nu >> 1$, the IPR approaches $1/N$.  We now focus on Fig.~\ref{fig:2}(b)-(c) where we show the IPR evolution at fixed shape parameter $\nu=100$ in the regime of time-homogeneous collisions. The IPR time evolution is displayed in Fig.~\ref{fig:2}(b) for various collision rates $r_c$ and fixed large disorder range $h=10$. We then show in Fig.~\ref{fig:2}(c) the IPR versus time for different disorder range constants $h$ from 0.1 to 10 at fixed low collision rate $r_c=0.1$. In the case of collisions that are homogeneous in time and in space, that is e.g. with $r_c = 0.1$ and $\nu = 100$ in the figure, noise appears in well-defined time slots. This makes visible the corresponding moments where magnetization and IPR change, exhibiting formation of plateaus in the IPR. In this case, we highlight that the higher the value of the disorder, the sharper the plateau. The plateaus are observed for high values of the shape parameter, e.g. while collisions occur at regular intervals, since noise is introduced into the system at well-defined times (see Fig.~\ref{fig:Model}(b)).  

For $r_c < 1$, localization followed by delocalization processes repeat iteratively for the excitation, leading the IPR to eventually approach its asymptotic value $1/N$, i.e. to the appearance of a fully delocalized state. We note that the cases with $\Delta \neq 0$ are not reported here because the inclusion of an anisotropic interaction term is seen to not qualitatively alter the system's behavior in the single-excitation regime, as expected considering that it contributes a boundary effect. For more details see \cite{AlessandroetAl}.

For $r_c \geq 1$ instead, the system is driven towards a diffusive regime. In fact, in this case the system transitions to a regime dominated by frequent collisions, which suppress the transient localization observed at lower collision rates. In this regime, the IPR evolves smoothly over time without forming plateaus, indicating sustained delocalization. High collision rates prevent the system from stabilizing in a localized state, continuously redistributing the excitation across the chain. However, we can say that for \corr{$r_c >5 $} the IPR increases with $r_c$, failing to reach his asymptotic value in our observation window. In fact, too high values of collision rate (in the figure $r_c= 50, 100$) slow down the dynamics and the spread of the excitation. This can be seen as a manifestation of the Zeno effect, as appeared also in the case of no disorder \cite{AlessandroetAl}.
\corr{This effect also causes the non-monotonic dependence on the collision rate which arises because collisions assist transport only when their rate is not too large. For $r_c \sim 5$ and above, frequent system-environment interactions act as repeated local projections that suppress coherent dynamics, slowing it, and finally freezing it, acting as strong disorder.} The overlap between the curves in Fig.~\ref{fig:2}(b) for $r_c = 0$ and $r_c = 0.1$ is to be expected (see also Fig.~\ref{fig:2_exc}(a) and Fig.~\ref{fig:EE}): indeed, a high shape parameter $\nu = 100$ causes the absence of collisions till $tJ = 10$, thus the system can be considered effectively closed, up to that time.

All in all, the comparison between homogeneous ($\nu \gg 1$) and heterogeneous ($\nu \leq 1 $) noise regimes reveals distinct transport behaviors. In fact, for homogeneous noise, collisions occur at regular intervals, leading to predictable dephasing effects and smoother transitions between localization and delocalization. In contrast, heterogeneous noise introduces temporal variability in collisions, resulting in intermittent localization and a more irregular evolution of the IPR. This highlights the role of noise structure in modulating transport dynamics.

We observe that noise generally leads to delocalization even in the presence of disorder for any collision rate $r_c \neq 0$. This behavior occurs because the noise-induced dephasing leads to progressive heating of the system, ultimately driving it toward a high-entropy $ T = \infty $ steady state with the excitation uniformly distributed across all sites. Disorder slows down the process by creating localized regions that act as traps and inhibits transport by delaying the redistribution of energy. Consequently, while the system eventually reaches the  $ T = \infty $ state regardless of the disorder strength, the time required to do so increases with higher levels of disorder. This highlights the competing roles of noise, which promotes delocalization, and disorder, which instead favors localization.

\subsection{\label{2_exc}Multiple excitations}

To shape our understanding of the system's dynamics, we now analyze the case of multiple excitations. Their interactions mediated by anisotropy, disorder and noise introduce additional complexity. We therefore switch to the Inverse Ergodicity Ratio (IER) described in Sec.~\ref{sec:Noise_des}~\cite{AlessandroetAl} and systematically analyze the interplay of the various parameters ($\nu$, $r_c$, $h$, $\Delta$) to map out plateau formation and delocalization scaling.

\subsubsection{\label{Twoexc}Two excitations case}

Let us step first on the case of two excitations injected in a chain subjected to noise with homogeneous collisions.

\begin{figure}[h!]
    \centering
    \includegraphics[scale = 0.6]{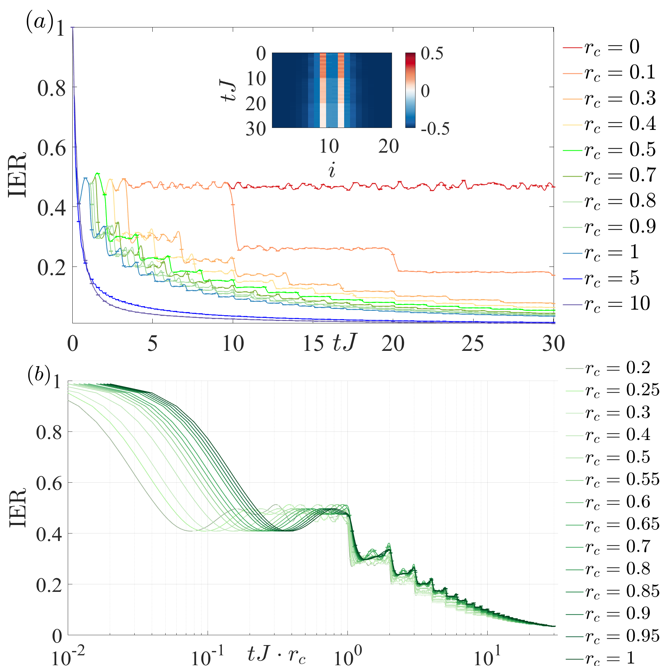}
    \caption{The case of two excitations. (a) Inverse Ergodicity Ratio (IER) vs time $tJ$ for different collision rates $r_c$ with anisotropy strength $\Delta = 2.5$. Inset: magnetization map over sites $i$ and time $tJ$, showing the spreading of the two excitations over time. The two neighboring excitations are initially separated by two spins with very homogeneous collisions over time ($\nu = 100$) and high disorder range with $h = 10$. The IER has the same trend of the IPR and we focus here on the collision rate regime $0 \leq r_c \leq 10$, proved to be more interesting due to the formation of plateaus. For lower collision rates ($r_c \ll 1$) the system remains localized (IER closer to 1), while increasing $r_c$ leads to a faster delocalization (IER decaying to its limit). (b) IER vs $tJ r_c$ varying the collision rate parameter $0 \leq r_c \leq 1$, at fixed high time-homogeneity ($\nu = 100$), high disorder strength ($h = 10$), and in the presence of anisotropy, considering two excitations initially separated by two spins. Curves with progressively increasing thickness refer to values of $r_c$ increasing from 0.20 to 1.00 by 0.05 each.  We see that all curves collapse in the region where plateaus are present since they are naturally shifted with respect to each other. The simulations were performed for a spin chain of $N=20$ sites with two excitations initially separated by two spins with $M = 250$ trajectories and timestep $dt = 0.02$ and final time of simulations $tJ =30$.} 
    \label{fig:2_exc}
\end{figure}

\begin{figure}
    \centering
    \includegraphics[scale = 0.22]{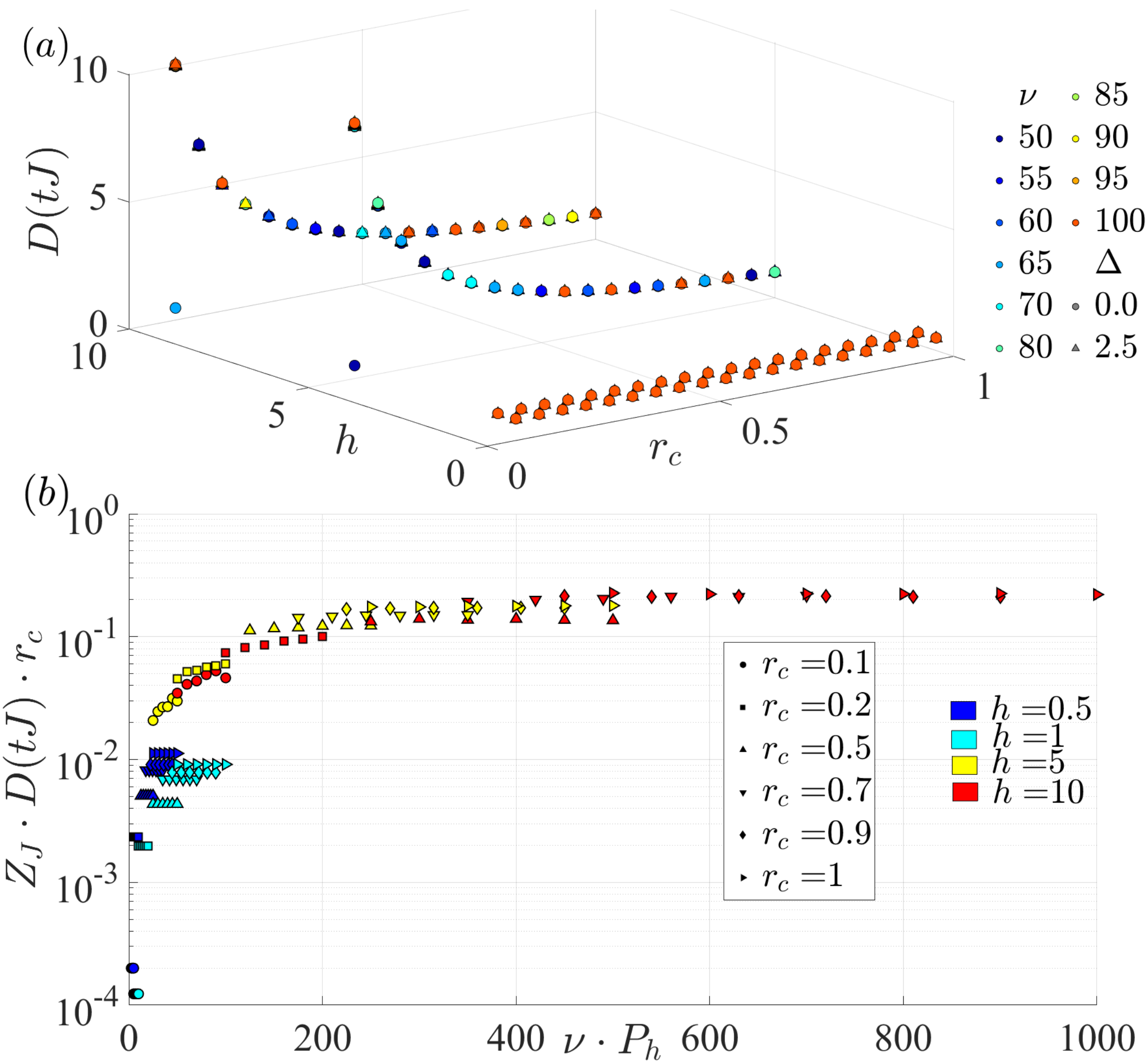}
    \caption{Characterization of the plateaus. (a) 3D plot of plateau \corr{duration} \corr{$D$} in terms of tJ vs disorder range $h \in [0,10]$ and  collision rate $r_c \in [0,1]$ (see Appendix \ref{2excmore}, Fig.~\ref{fig:2dplot}, for a 2D version). Different symbols colors and shapes distinguish different levels of collision time-homogeneity (colors) and anisotropy (shapes), as in the legend. (b) Representing all the relevant datasets with only two quantities that collect all the governing physical parameters. Semi-log plot of the dimensionless plateaus area  \corr{$Z_J D(tJ) r_c$} vs the disorder power per unit collision time $P_h\equiv hr_c$, scaled with the shape parameter $\nu$ (see text for a discussion). The datasets can be identified from the legend: different symbols label the collision rates $r_c = 0.1, 0.2, 0.5, 0.7, 0.9,1$, different colors the disorder strength values $h = 0.5, 1, 5, 10$. Notice that all the datasets collapse in two curves: in essence, no plateaus for low disorder for any other parameter (bluish symbols) and rapidly developing plateaus to a saturating value with increasing disorder power, for slow collision rates and any other parameter (yellow-reddish symbols). As already commented, the anisotropy $\Delta$ does not qualitative impact the results and it is therefore fixed at $\Delta = 2.5$ (that is, other $\Delta$ values would be not distinguishable). Also, adding the data with $r_c > 1$ would result in having an horizontal lineup of data points collapsed at vanishing plateaus area irrespective of all the other parameters, and are thus not reported. The simulations were performed for a spin chain of $N=20$ sites with two excitations initially separated by two spins, with $M = 250$ trajectories, timestep $dt = 0.02$, and final time of simulations $tJ =30$.}
    \label{fig:2_exc_2}
\end{figure}

\begin{figure}
    \centering
    \includegraphics[scale = 0.145]{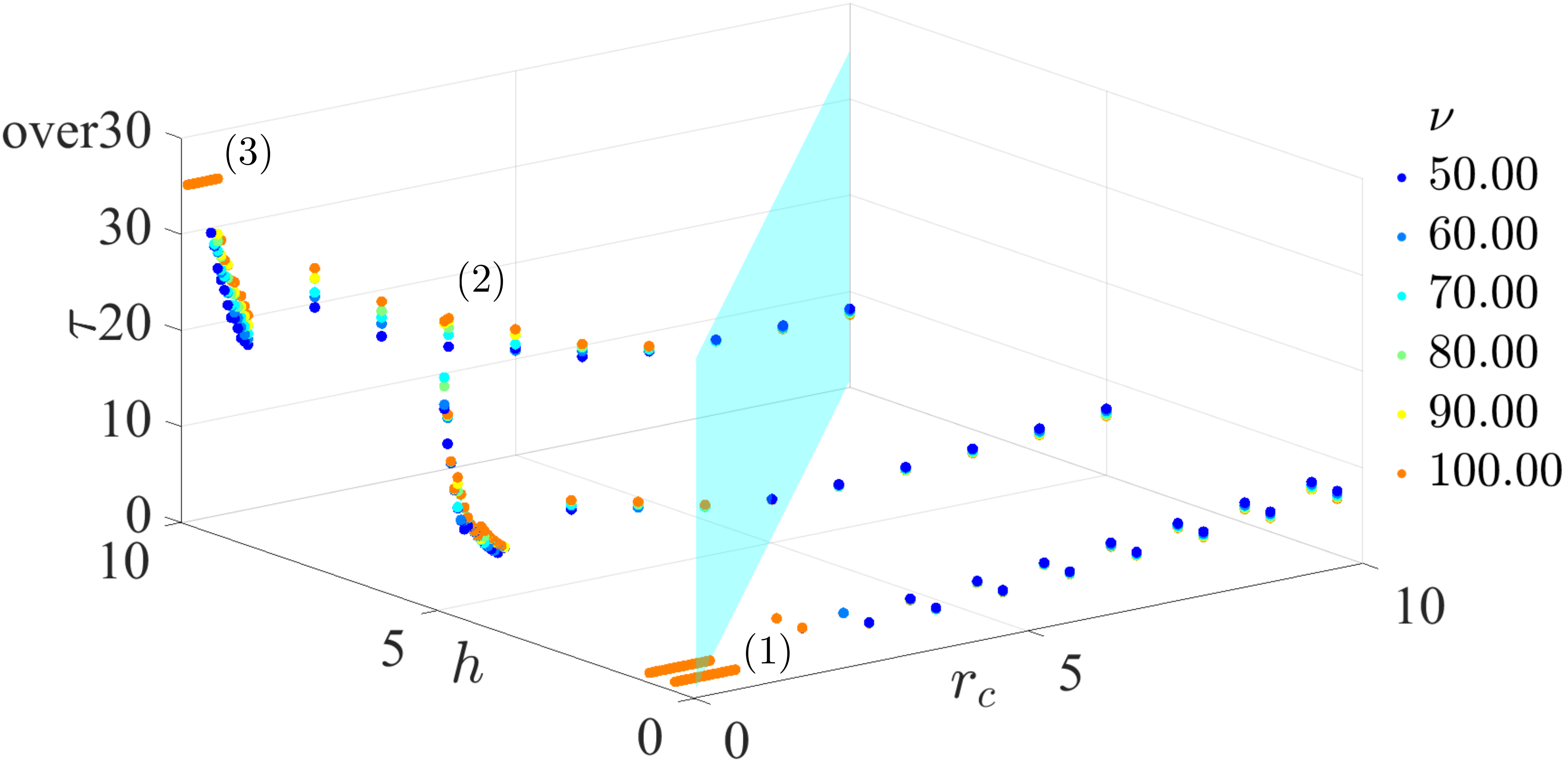}
    \caption{Study of delocalization time $\tau$ (see text, Sec.~\ref{2_exc}). 3D plot of the complete delocalization time $\tau$  vs $h \in [0,10]$ and $r_c \in [0,10]$. The plane $r_c = h$ separates small from intermediate-to-high $\tau$ regions. Region (1) with small $\tau$ does not exhibit plateaus, while the region with plateaus occurs in two regimes: with intermediate (2) and long (3) $\tau$, corresponding to a more localized/delocalized behavior (see text, Sec.~\ref{2_exc}). Symbols with different colors represent different levels of collision time-homogeneity as in the legend. The simulations were performed for a spin chain of $N=20$ sites with two excitations initially separated by two spins with $M = 250$ trajectories and timestep $dt = 0.02$ and final time of simulations $tJ =30$. The data with $\tau > 30$ are those for which the system does not succeed to reach complete delocalization in that time. }
    \label{fig:2exc_3}
\end{figure}

We show in Fig.~\ref{fig:2_exc}(a) the time evolution of the IER for two excitations initially injected at a distance of two spins apart in a system of $N=20$ sites, subjected to collision rates $r_c$ (from $0$ to $10$) at fixed temporal homogeneity $\nu = 100$, anisotropy $\Delta = 2.5$, and a high disorder range $h = 10$. The general trend indicates that lower collision rates ($r_c \ll 1$) lead to prolonged localization with the IER remaining closer to 1: the system's state is dominated by a limited number of basis configurations. 
As $r_c$ increases, the IER decreases more rapidly over time, signaling enhanced delocalization due to frequent noise-induced perturbations. For $r_c \sim 0.5$, we observe a gradual reduction of the IER, while for higher rates $r_c \geq 1$, the system rapidly approaches ergodic behavior.
For completeness, we display in the inset of Fig.~\ref{fig:2_exc}(a),  the magnetization spread at fixed collision rate $r_c = 0.1$, the other parameters being unchanged.

Consistently with the single-excitation scenario, the emergence of plateaus in the IER remains evident in the presence of homogeneous noise with a low collision rate. The presence of anisotropy does not modify the IER trend when the two excitations start separated by two spins, thus in this case can be considered a sub-leading effect. However, the plateaus are stabilized by the anisotropy interaction term even for lower disorder (small $h$ values) when the excitations are injected next to each other (see the difference between Fig.~\ref{fig_app:2_exc}(a) and Fig.~\ref{fig_app:2_exc}(b) in Appendix B).  This suggests that when more excitations are present, anisotropy works to favor the system localization. 

In contrast with the case of one excitation, an increase in the collision rate ($r_c = 1,5,10$) promotes transport by breaking the pinning of the excitations, an effect facilitated by noise. In fact, the presence of more than one excitation leads to a rapid decrease of the IER for higher values of the collision rate. This can be seen as a manifestation of
stochastic resonance, in fact happening also without disorder \cite{AlessandroetAl}. 

We notice from Fig.~\ref{fig:2_exc}(a) that, before the plateaus are well developed, the IER curves at different $r_c$ values tend to collapse, indicating indeed that delocalization gets progressively insensitive to collision events once these become more frequent. It is therefore natural to ask whether a universal-type behavior can be identified in the regime of rare collision events. To this aim, we show the IER vs. time rescaled with $r_c$ for two excitations initially separated by two spins in the same spin chain with $N=20$ sites, in the case of anisotropy $\Delta =2.5$, high shape parameter $\nu = 100$, high disorder range $h=10$, and varying the collision rate $r_c$ between 0.2 and 1  (Fig.~\ref{fig:2_exc}(b)).  We indeed see that all curves collapse in the region where plateaus are present. This occurs because the plateau \corr{duration} scales as $1/r_c$ and the different curves are thus naturally shifted with respect to each other.

To have a complete understanding of the plateaus formation, we have systematically explored how all the governing system parameters (disorder, time and space homogeneity, and anisotropy) affect the degree of localization.

To this end, we introduce the \corr{duration} of the plateaus \corr{$D(tJ)$} and the jump size $Z_J$. \corr{$D(tJ)$} is the duration of the first plateau for each curve along the x-axis in units of $tJ$. $Z_J$ is the height of the first plateau for each curve.

In Fig.~\ref{fig:2_exc_2}(a), we show the 3D plot of \corr{$D(tJ)$} as a function of the collision rate $r_c$ and the disorder range constant $h$, with the shape parameter $\nu$ and the anisotropy $\Delta$ varied as in the legend. We can see here that the data with different values of anisotropy collapse, thus from now on in this section we will present only data with $\Delta = 2.5$. (For a differently expressive 2D version of the plot, we refer to  Appendix \ref{2excmore}, Fig.~\ref{fig:2dplot} \corr{(a)}). To make order in all the data, we identify physically meaningful quantities that can collect all the governing system parameters. To start with, we consider the plateaus area, that is their height, i.e. the jump size $Z_J$, times their dimensionless \corr{duration} \corr{$D(tJ)r_c$}: this quantity can be viewed as a sort of order parameter for the appearance of the plateaus. We then collect the remaining physical quantities in $hr_c\nu$, the rate of change of disorder bandwidth $h$ per unit collision time $r_c$, scaled with the shape parameter $\nu$: in fact, $P_h\equiv h \cdot r_c$ can be viewed as form of disorder power in collision-time units. We present in Fig.~\ref{fig:2_exc_2}(b) a semi-log plot of \corr{$Z_JD(tJ)r_c$} vs. $P_h\nu\equiv h r_c\nu$. The remaining parameters are as usual: $N=20$ sites for the spin chain, two excitations initially separated by two spins, anisotropy $\Delta = 2.5$, collision rates $r_c$ in the range between 0.1 and 1 (labeled by different shapes as in the legend), different disorder-range values $h = 0.5,1,5,10$ (shown with different colors as in the legend), and different values for the shape parameter $\nu$ in the homogeneous range between 50 and 100. We see that all the data collapse in essentially two curves, distinguishable at a glance by bluish and yellow-reddish symbols. In fact, we can easily identify the region corresponding to no plateaus, i.e. with vanishing plateaus area: this corresponds to low values of disorder strength (0.5 and 1, blue and light-blue dots) and to high collision rates $r_c > 1$ interdependently of all the other parameters. For larger values of $h> 1$, instead, the plateaus rapidly develop with decreasing the collision rate $r_c$, up to a size that saturates to a value of $\approx 0.2$ independently of the other parameters. While this value could change with the number of excitations involved, it would be interesting to explore the degree to which it might have a universal meaning. 

Before proceeding further, let us summarize our findings so far. Delocalization is facilitated by conditions that minimize the stabilizing effects of disorder and anisotropy. Low disorder levels ($h \ll 10$) and weak anisotropy allow noise-induced scattering to overcome any tendency toward localization. This occurs especially when combined with a higher collision rate ($1 \leq r_c \leq 10$), which disrupts coherent trapping mechanisms. Furthermore, lower values of the shape parameter $\nu \sim 50$ contribute to more dynamic and irregular scattering events, also promoting delocalization. In contrast, higher levels of disorder ($h \geq 10$) and anisotropy, coupled with lower collision rates ($r_c \ll 1$), result in a pronounced localization. In this regime, disorder-induced traps dominate the dynamics, leading to extended plateaus where the system remains effectively localized over long timescales. We observe that the \corr{duration} of these plateaus scales inversely with the collision rate ($\sim 1/r_c$), consistently with the one excitation case: indeed, low collision frequencies allow the system to persist in localized configurations for extended periods of time, before noise-induced delocalization sets in.

To gain further insight on the system delocalization behavior, we now introduce the complete delocalization time $\tau$. This is defined as the first time (in terms of tJ) at which the IER falls below $10/dimH$ (10 times its lower bound), and after which no additional plateaus are clearly distinguishable. Our findings summarized in Fig.~\ref{fig:2exc_3} highlight an intricate interplay in shaping the localization dynamics.  
In Fig.~\ref{fig:2exc_3} we show $\tau$ in a 3D plot  again versus disorder range $h$ and  collision rate $r_c$, and for different values of time-homogeneous shape parameters $\nu$ and anisotropy $\Delta = 2.5$. \corr{(For a differently expressive 2D version of the plot, we refer to  Appendix \ref{2excmore}, Fig.~\ref{fig:2dplot} (b))}

We again identify two regions: small and intermediate-to-long $\tau$. The first region (1) refers to a complete delocalization time $\tau \leq 10$, here happening for two datasets: data with low collision rates $0.1 \leq r_c \leq 1$ combined with small level of disorder $h \leq 1$ and data with higher collision rates $r_c > 1$. In all cases, $\tau$ is small independently of the shape parameter $\nu$ and no plateaus are sizable. In the second region pleateaus build up but in two different regimes, (2) and (3). In (2), complete delocalization occurs over a time $10 \le \tau \leq 30$, happening for higher level of disorder $h = 5, 10$ combined with an intermediate collision rate $1 \leq r_c \leq 5$. Finally, in (3) complete delocalization occurs over time $\tau > 30$, occurring for high disorder with $h=10$, high collision time homogeneity $\nu =100$, and small collision rates $r_c \leq 1$. Notice that this region, not determined in our observation window for the IER has not reached its lower bound, corresponds to the formation of the longest plateaus. 
Using the plane $r_c = h$ as a reference, we observe that for $r_c < h$, the complete delocalization time remains high, while for $r_c > h$, $\tau$ is significantly reduced. This behavior resembles the case without disorder (\cite{AlessandroetAl}), where the complete delocalization time was indeed limited to $\tau \leq 5$. 

Summing up, we can conclude how delocalization turns out to be facilitated by low levels of disorder, anisotropy, and time homogeneity, along with a higher collision rate ($ 1 < r_c < 10$). Instead, higher levels of disorder and anisotropy, coupled with a lower collision rate, lead to more localized behavior, with the size of the plateaus scaling as $1/r_c$. 

\subsubsection{\label{Mult_exc}The full picture}

Having framed the most interesting parameter regions, we are now ready to get the full picture \last{by exploring}  the more general case of multiple excitations. 

To this aim, for computational reasons we shrink our spin chain to $N=8$ and inject $N_{exc} = 4$ into the system. We initially inject them all next to each other, as we have seen that this is the condition to size the effect of anisotropy. We also restrict the parameter range to the most significant behavior, as learned from Sec.~\ref{Twoexc}. To sum up, this is for  space-homogeneous (e.g. $\nu = 100$) collisions happening at smaller rates,  high disorder range bandwidth (e.g. $h=10$) and anisotropy as large as $\Delta = 2.5$. 

\begin{figure}
   \centering
   \includegraphics[scale = 0.15]{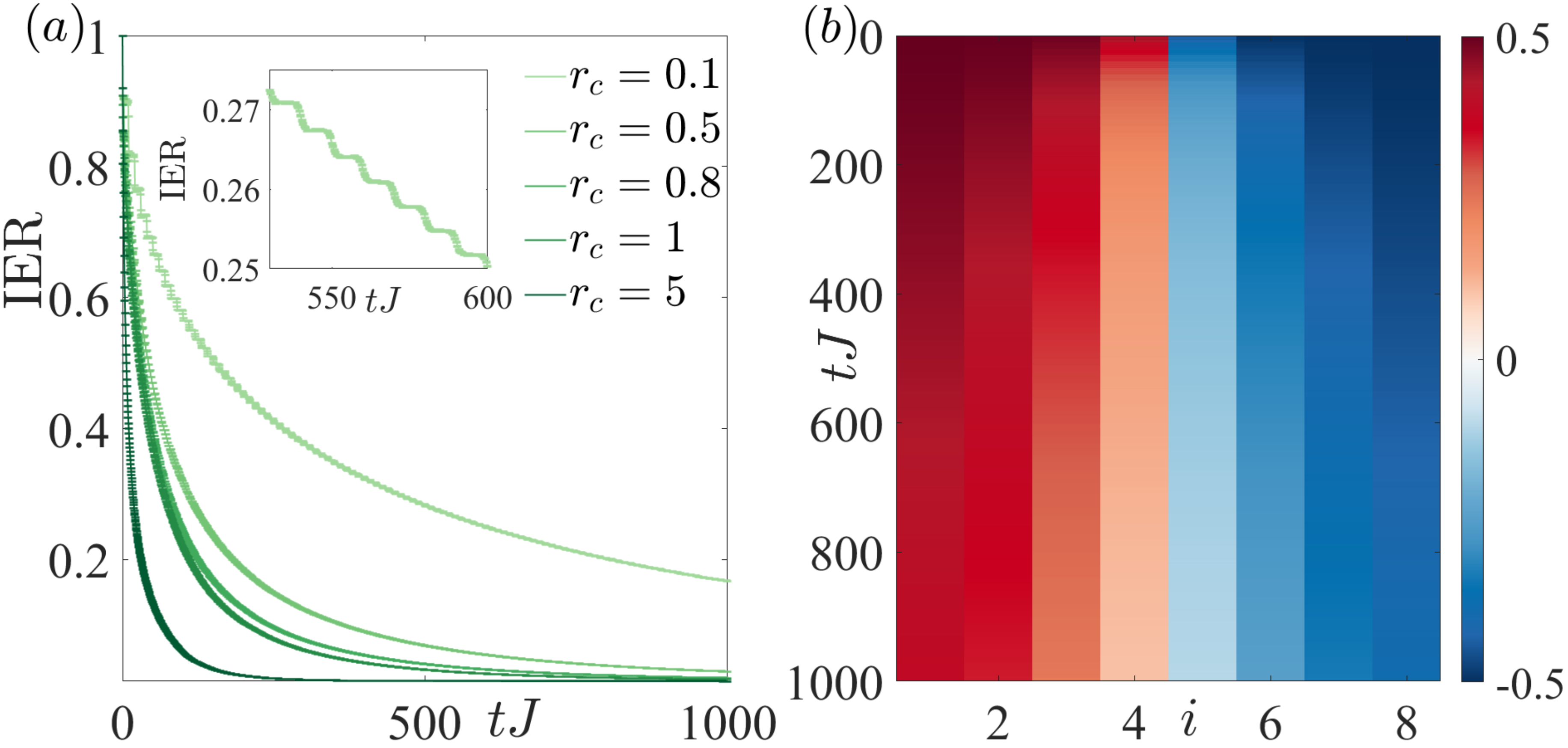}
    \caption{Case of multiple excitations. Inverse Ergodicity Ratio (IER) (a) and color plot of the magnetization (b) for a spin chain of $N=8$ sites and $N_{exc} = 4$ excitations starting \last{ on the left side of the chain, producing a domain-wall initial state}, time-homogeneous collisions $\nu = 100$, with high disorder range width $h=10$ and anisotropy strength $\Delta = 2.5$. (a) IER vs time $tJ$ for different collision rates $r_c$ in the interesting regime as in the legend. Inset: Zoom of Fig.(a). (b) Spread of magnetization over time $tJ$ and sites $i$ in the interesting regime of low collision rates, here e.g. with $r_c= 0.1$. We can overall see that low collision rates ($r_c \ll 1$) combined with time-homogeneous collisions ($\nu = 100$) lead the system to remain localized for a significant time, prolonged also by the presence of anisotropy and high disorder. Only at very long times the IER approaches its asymptotic value, signaling full delocalization. The simulations were performed with $M = 250$ trajectories and timestep $dt = 0.02$. Final time is $tJ =1000$. }
    \label{fig:multiple_exc}
\end{figure}

We display in Fig.~\ref{fig:multiple_exc}(a) the time evolution of the Inverse Ergodicity Ratio (IER) up to enough long time $tJ = 1000$, to let the system evolve and potentially reach a delocalized state. We see that for low collision rates ($r_c \ll 1$) combined with time-homogeneous collisions, ($\nu \gg 1$), the system struggles to delocalize. In such conditions, the interplay of disorder, anisotropy, and noise stabilizes localized states delaying the onset of delocalization as it is zoomed in the inset. In this case, the height of the plateaus is shorter compared to the case of one and two excitations and complete delocalization is more difficult to reach. Thus, only at very late times does the IER approach its asymptotic value, indicating that the system has completely delocalized.

To gain insight on the underlying spatial behavior, we look in Fig.~\ref{fig:multiple_exc}(b) ata color plot for the spread of magnetization over time. We see that for low collision rates and high time-homogeneous collisions, the magnetization piles up for extended periods of time in specific regions of the spin chain, highlighting the system's localized nature. Anisotropy and disorder favor the pinning of excitations to certain sites. As time progresses, however, collisions eventually disrupt these pinned configurations, leading to gradual spread of magnetization across the chain. Delocalization is however significantly delayed under these conditions, requiring long timescales for the system to be fully delocalized.

Before complementing the dynamics characterization with the information that can be extracted from the entanglement entropy, we conclude this section by summarizing in Table~I the behaviors described so far.\\ 
\begin{table}[]
\begin{tabular}{|l|l|l|} 
\hline
Parameter & Delocalization & Localization \\
\hline
Collision rate & $r_c > 1 $ & $r_c \leq 1$ \\
\hline 
Shape parameter & small $\nu$ & $\nu >> 1$ \\
\hline
Disorder strength & $h = 0.5, 1$ & $h = 5, 10$  \\
\hline
\end{tabular}\label{Tab1}\\
\noindent
\caption{ The full picture. Summary of the localization/delocalization regimes varying the collision rate, the shape parameter and the disorder strength.}
\end{table}

The table summary is completed by reminding how time non-homogeneous collisions (low $\nu$) and varying anisotropy $\Delta$ influence the dynamics. If we look at regime (3) in Fig.~\ref{fig:2exc_3}, we notice that the data with small collision rate, high disorder strength and $\nu = 100$ have the highest complete delocalization time $\tau$. Thus, in general very high value of $\nu$ favors the presence of localized regions. However, when the collision rate increases the shape parameter plays a sub-leading role in localizing/delocalizing the excitations.
Also, we notice that the anisotropy $\Delta$ plays a more important role for multiple excitations, enhancing localization when the excitations are next to each other (see also Appendix \ref{2excmore}).

\subsection{\label{EE}Entanglement Entropy}

To complement the studies based on local observables, in this section we analyze the behavior of entanglement as a global measure of thermalization dynamics in our spin chain. We focus on the case of two excitations injected into the system next to each other, where disorder, noise and coupling between spins significantly affect the delocalization behavior. We limit to two excitations for computational ease, having confirmed that the multiple excitations case has the same trend as the two excitations case in Sec.~\ref{2_exc}. Finally, we again zoom the parameters in the interesting range emerged in Sec.~\ref{Twoexc} with the emergence of plateaus and non-continuous temporal profiles of transport: this corresponds to low collision rates, e.g. with $r_c = 0.1$, large disorder range e.g. with $h =10$, and important anisotropy e.g. with $\Delta = 2.5$. Finally, we now want to check the dependence on the noise space-homogeneity by letting the shape parameter $\nu$ vary from absence of noise ($\nu = 0$), to heterogeneous collisions over time ($\nu \leq 1$) and time-homogeneous collisions ($\nu = 100$).  

\begin{figure} [h]
    \centering
 \includegraphics[scale= 0.15]{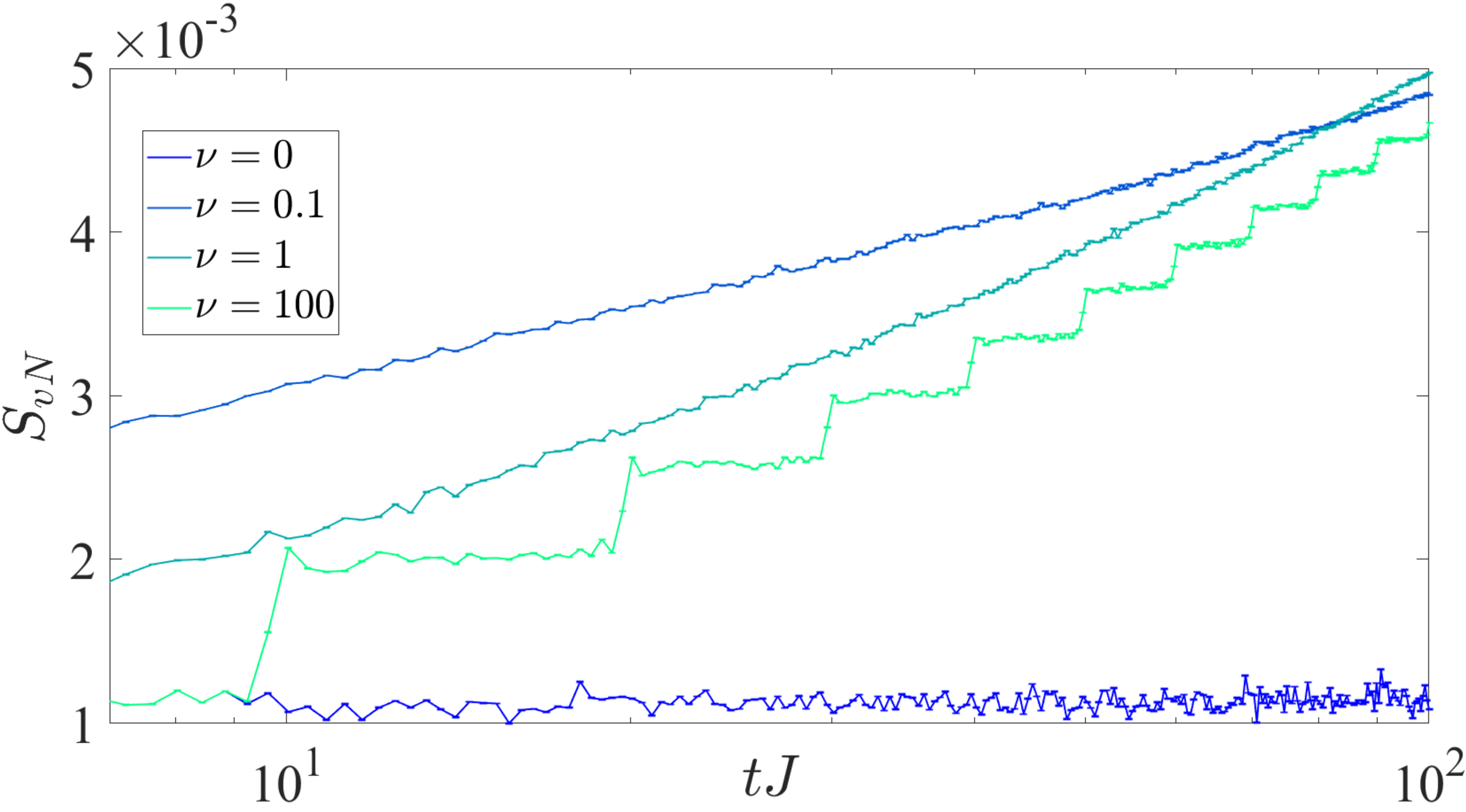}
    \caption{Entanglement Entropy and dependence on noise space heterogeneity. $S_{vN}$ vs time $tJ$ in a semi-log scale for different levels of collision-time homogeneity $\nu$ and fixed low collision rate $r_c = 0.1$, large disorder range $h =10$ and anisotropy $\Delta = 2.5$ in the interesting plateaus regime, for a disordered spin chain of $N = 20$ sites and two injected excitations starting next to each other. The simulations were performed with $M = 250$ trajectories and timestep $dt = 0.02$.}
    \label{fig:EE}
\end{figure}

We then evaluate the Von Neumann entropy of the corresponding reduced density and present in Fig.~\ref{fig:EE} the time evolution of the entanglement entropy $S_{vN}(t)$ with the bipartition 10-19 in a semi-log scale. 

In general, we expect that the low density of excitations limits the number of degrees of freedom available for generating entanglement: fewer interacting particles result in reduced correlations. Furthermore, we have moderate transport of information due to the presence of disorder, anisotropy and noise limiting the spread of quantum correlations throughout the system. 
In the absence of noise $\nu= 0$, we observe in Fig.~\ref{fig:EE} a flat line over time, apart from some fluctuations. In fact, without noise driving the diffusion of correlations over the spin chain, the system persists in a coherent and localized state dominated only by the effects of disorder and anisotropy. \corr{For values of $\nu \neq 0$, we observe a slow logarithmic growth of $S_{vN}$, consistent with the breaking of localization due to collisions \cite{Abanin2019,Wybo,Lukin,Fischer}. In the homogeneous noise regime ($\nu =100$), the system exhibits a slow growth between successive plateaus. Instead, when $\nu$ is smaller, we see an accelerated
breaking, resulting in a faster growth of $S_{vN}$. This is due to the introduction of more collisions, being the effective collision rate $r_c$ higher in this case.}  Even with increasing collision  time-homogeneous stroboscopic noise ($\nu =100$), we observe the ability to also disrupt localization with the formation of plateaus. While delocalization is slower, we can tailor its rate by modifying the timescales of the plateaus as discussed in the previous sections.

\section{\label{Discc}Discussion\\}

In this study, we analyze how coherent and dissipative couplings can manipulate the transport properties and thermalization of a noisy, disordered XXZ spin chain, where stochastic collisional noise simulates the interactions with its environment. We identify relevant regimes in the rich landscape determined by the interplay between disorder, anisotropy, and collisional noise to shape the localization/delocalization transitions and the specific transport rates. 

The impact of this work is twofold. Firstly, in analyzing our datasets we could extract forms of scaling behaviors, as e.g. for the appearance of plateaus (see Figs. ~\ref{fig:2_exc} and ~\ref{fig:2_exc_2}): in so doing, we make order in the plethora of data emerging from our systematic study. These scaling behaviors represent valuable guidelines to think of the system. Further studies under more complicated yet interesting conditions, such as non-conserved magnetization can follow, having our present work as a benchmark. Secondly, we have worked our results out using the paradigmatic XXZ model as playground and an enough general, two-parameters, stochastic model to shape the collisions besides the more conventional addition of disorder: in this perspective, our findings can have many potential applications in specific systems, qualitatively and quantitatively.  

Turning to the specific findings, from the simplest one excitation case, see Sec.~\ref{One_Exc}, we found that disorder tends to suppress transport through interference: in so doing, it competes with collisional noise, that instead facilitates delocalization through dephasing. One first message emerging from our study is that among these competing mechanisms, low collision rates ($r_c \ll 1$) favor the occurrence of localized regions in the form of plateaus for the Inverse Participation Ratio (IPR), in time windows where collisions are absent. Interestingly, the area of these plateaus depend directly on noise parameters with a form of universal scaling towards a saturating value. The degree of universality for this behavior certainly requires further investigation. Overall, the occurrence of these plateaus allows to systematically tailor sequences of transport and localization that can be applied for example to stroboscopic protocols in quantum technologies. At higher collision rates ($r_c \geq 1$) instead, frequent collisional events completely disrupt localization, driving the system toward a diffusive regime characterized by smooth evolution of IPR over time, even for moderate levels of disorder range. The presence of these localization/delocalization transitions can be understood for single excitations, since collisions do not alter the total magnetization and the system remains localized between collisions, preventing the IPR from decaying during these intervals.

From the case of several excitations, see Sec.~\ref{2_exc} and Appendix B, we learn that additional features emerge due to interactions among excitations. In particular, more localized regions in the form of plateaus appear even for the tiniest levels of disorder range. However, we still find that increasing collision rates ($r_c > 1$) disrupt localized states, promoting transport. We demonstrate the role of noise in overcoming pinning effects, down to complete delocalization in a short time $\tau$. This is similar to what we found in the absence of disorder \cite{AlessandroetAl}, where the complete delocalization times did not exceed $tJ = 5$. In addition, we also find regimes of tunable transport and the presence of plateaus, in fact even at higher densities. \corr{In regimes of high disorder and anisotropy, collisions induce a gradual logarithmic entanglement growth. This can be understood as a signature of the progressive breaking of the localized regime, rather than its preservation, consistent with the presence of localized states that prevent rapid spread of information.} For collisions heterogeneous in time instead, the entanglement entropy more rapidly grows. Thus, homogeneous noise leads to smoother entropy dynamics.

Our analysis is of relevance for quantum technologies applications. For instance, in quantum circuits optimizing transport is fundamental, for the efficiency of information flow directly impacts computational performance \cite{Quantum_circuit}. In fact, besides using thermodynamic benchmarks such as ergotropy, power output, and efficiency, energy devices like quantum batteries and quantum heat engines could benefit from noise-engineering to enhance their performance in terms of efficiency and stability~\cite{Carrega2020, Mayo2022}. Also, the precise manipulation of spin transport in systems such as NV centers in diamond has already enabled ultra-sensitive magnetic field measurements for quantum sensors and processors~\cite{Jelezko2012, Oberg2019}. In addition, understanding how localization-delocalization transitions arise in the presence of noise could help designing novel materials with tailored transport properties \cite{Nanowires}. 

Noise-disorder engineering could also be used to optimize energy transport in artificial photosynthetic complexes, in essence using controlled noise to enhance excitonic energy transfer, and mimicking the noise-assisted transport observed in natural photosynthesis. In addition, tailoring noise properties at a microscopic scale in biological systems, where energy transfer is influenced by disorder and environmental fluctuations,  could help to balance localization and delocalization effects, aiming at maximizing efficiency.
Under an alternative perspective, this offers means to control energy flow in quantum networks, by tuning the connectivity and interaction strength between nodes. Along these lines, the development of programmable quantum devices could be facilitated where energy or quantum information is routed dynamically. In fact, recent works apply reinforcement learning and neural-network optimization to tailor quantum channels for target tasks and evolutionary algorithms can optimize control pulses for population transfer in multilevel systems. These techniques can be used to design collisional sequences that maximize transport efficiency or plateau tunability \cite{Machine_Learn_25}.

Finally, one more interesting perspective is to move beyond purely Markovian collisional models, considering that time-correlated environments can qualitatively alter transport. For instance, Random Telegraph Noise (RTN) and Ornstein–Uhlenbeck Noise (OUN) preserve coherence bursts that enable high-fidelity state transfer even on complex graphs \cite{No_markov_2025}. Embedding such correlated noise kernels into our collision model could reveal richer localization–delocalization transitions.

\begin{acknowledgments}
The authors would like to thank A. Daley, G. M. Cicchini, M. C. Morrone, L. Guidi, C. Ceccanti, M.Landi, C.V. Stanzione and S. Ausilio
for inspiring and useful discussions. V.S. and M.L.C. acknowledge support from the National Centre on HPC, Big Data and Quantum Computing - SPOKE 10 (Quantum Computing) and received funding from the European Union Next-GenerationEU - National Recovery and Resilience Plan (NRRP) MISSION 4 COMPONENT 2, INVESTMENT N. 1.4 CUP - N. I53C22000690001. J.Y.M. and M.L.C. were supported by the European Social Fund REACT EU through the Italian national program PON 2014-2020, DM MUR 1062/2021. M.L.C. also acknowledges support from
the project PRA2022202398 ``IMAGINATION''.
\end{acknowledgments}

\appendix

\section{ \label{App_Imb}Imbalance}

In addition to IPR and IER, in the presence of disorder, the imbalance IMB provides a descriptive figure of merit of the system's relaxation dynamics. 
Indeed, we can write this figure of merit as:
\begin{equation}
    IMB = n_c(t) - \tilde{n}(t) ,
\end{equation}
\corr{where $n_c(t) =1 / 2\big(1+\langle \sigma_c^z(t)\rangle\big)$ is the local density of excitations at the central site of the chain, and $\tilde{n}(t) = {n(t)}/{N}$, with
   $ n(t) = \sum_{i=1}^N 1/2 \big(1+\langle \sigma_i^z(t)\rangle\big)$ ,
and $N$ the total number of sites.}

This definition incorporates the requirement that at time $t=0$ the excitations are only present on the central site, \last{$n_c(0) = 1$, so that $\tilde{n}(0) = N^{-1}$ } and imbalance \last{$IMB = 1 - N^{-1}$}.  After letting the system evolve, the excitation will spread through the system to a certain degree depending on the parameters. In a \improv{fully} delocalized regime we have $n_c = 1/N$ and $\tilde{n} = 1/N$, leading to $IMB = 0$.  

\begin{figure} [h!]
    \centering
    \includegraphics[scale= 0.285]{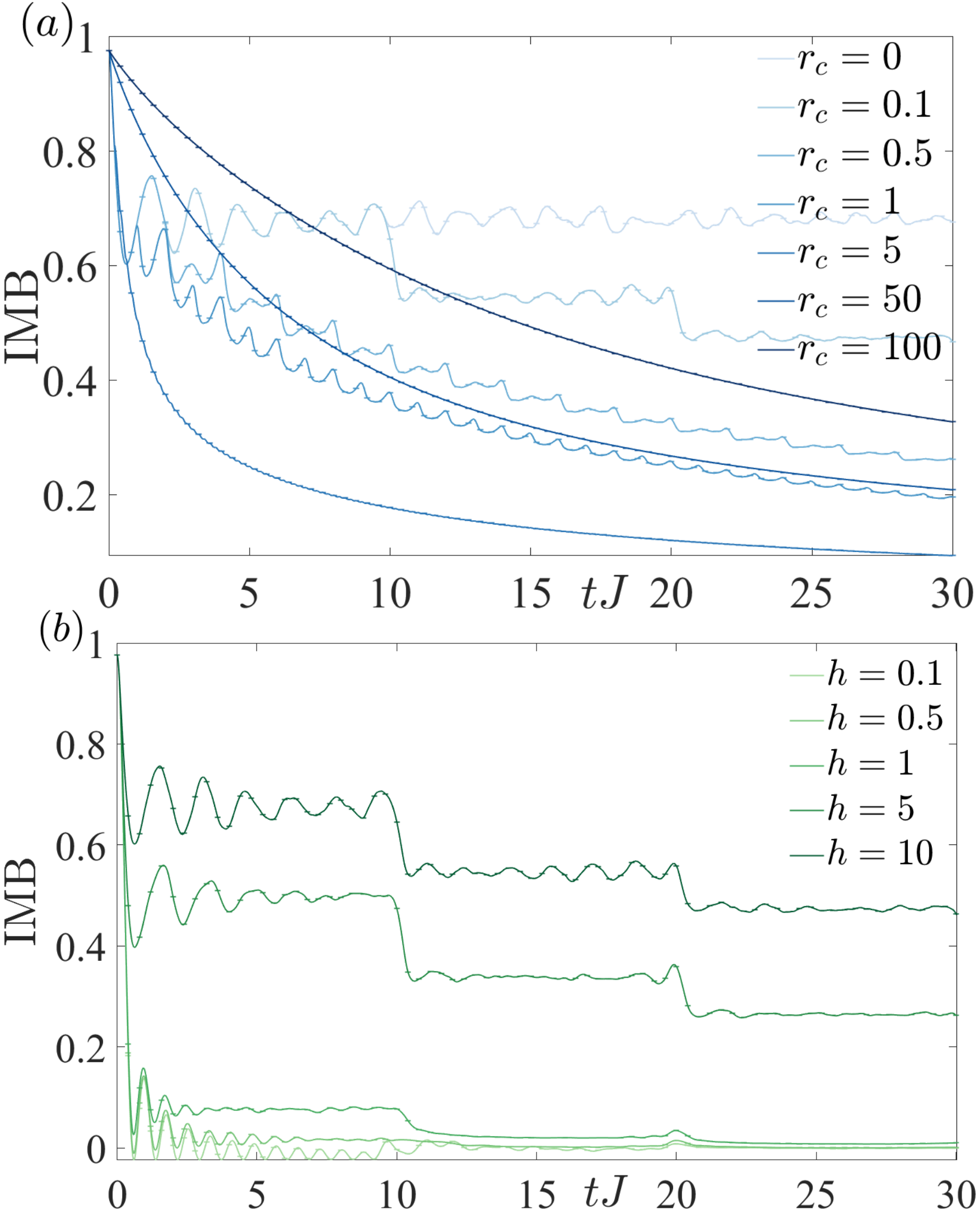}
    \caption{Imbalance for the One excitation case. (a) Imbalance for a spin chain of $N=41$ sites and $N_{exc} = 1$ excitations starting in the middle of the chain, shape parameter $\nu = 100$, disorder range constant $h=10$ and interaction strength $\Delta = 0$ for different collision rates. Final time is $tJ =30$. (b) Imbalance versus time for a spin chain of $N=41$ sites, one excitation, different disorder range constants at fixed collision rate $r_c=0.1$, shape parameter $\nu = 100$ and anisotropy $\Delta = 0$.  The simulations were performed with $M = 500$ trajectories and timestep $dt = 0.02$.}
    \label{fig:A1}
\end{figure}

In Fig.~\ref{fig:A1}, we analyze the time evolution of the imbalance $IMB$ for a disordered spin chain with $N = 41$ sites, one excitation initially localized at the middle of the chain, and shape parameter $\nu = 100$. In particular, Fig.~\ref{fig:A1}(a) shows how the Imbalance behavior varies over time for different collision rates, while Fig.~\ref{fig:A1}(b) shows its dependence on different disorder \improv{strengths} at fixed $r_c = 1$. The imbalance $IMB$ quantifies the difference in population between the initially occupied and unoccupied regions of the chain, providing insight into the dynamics of localization and delocalization. For low collision rates ($r_c \ll 1$), the imbalance decays very slowly over time, reflecting strong localization induced by the combined effects of disorder and anisotropy.  

As the collision rate increases, noise events become more frequent, leading to a faster decay of $IMB$. This behavior indicates improved delocalization, where excitation spreads more uniformly across the chain, erasing the initial population imbalance. However, as \improv{with the IPR}, higher values of $r_c$ (50, 100) \improv{result} in a slowdown of the dynamics of the system, manifestation of the \improv{quantum} Zeno effect \cite{AlessandroetAl}.

\section{\label{2excmore} More data on the two excitations case}

\begin{figure}[h!]
    \centering
    \includegraphics[scale = 0.285]{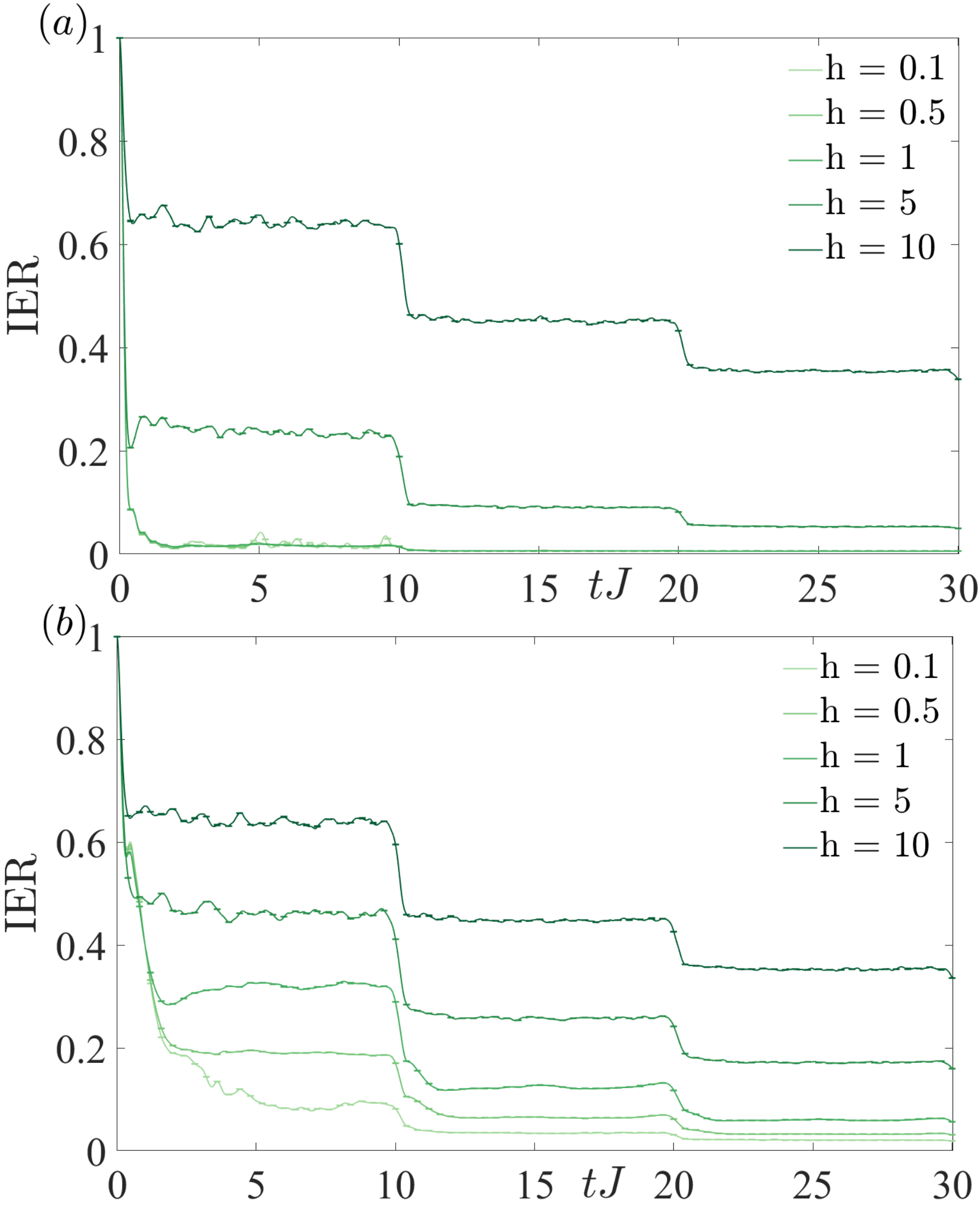}
    \caption{Two excitations case. Inverse Ergodicity ratio (IER) for a spin chain of $N = 20$ sites and final time of simulations $tJ = 30$. (a) IER vs time $tJ$ for different disorder range constants at fixed $r_c = 0.1$, $\nu = 100$, $\Delta = 2.5$ for two excitations starting initially separated by two spins and (b) two excitations starting next to each other. The simulations were performed with $M = 250$ trajectories and timestep $dt = 0.02$.}
    \label{fig_app:2_exc}
\end{figure}

Here, we show additional results for the case of two excitations injected into the disordered spin chain. In particular, we present what happens when anisotropy plays a more fundamental role in the dynamics of the system, by analyzing the Inverse Ergodicity Ratio (IER) under varying disorder regimes and excitation configurations. For excitations starting next to each other (Fig.~\ref{fig_app:2_exc}(b)), anisotropy amplifies the formation of plateaus in the IER, even for low disorder levels. This can be seen as a more localized behavior in the presence of a major potential energy due to the proximity of the excitations, leading to a minor kinetic energy.  In contrast, when the excitations are initially separated (Fig.~\ref{fig_app:2_exc}(a)), the effects of anisotropy are subdominant, with the dynamics primarily governed by disorder and noise. These results reinforce the particular interplay between noise, disorder, and anisotropy in modulating transport properties, offering further insights into the mechanisms driving localization and delocalization transitions. The plots are for a spin chain of $N = 20$ sites, at fixed $r_c = 0.1$, $\nu = 100$, $\Delta = 2.5$, varying the disorder range constants between 0.1 (low disorder) and 10 (high disorder). We report here also \corr{a semi-log} 2D version of the 3D plot in Sec.~\ref{2_exc} Fig.~\ref{fig:2_exc_2}. \corr{Focusing on} higher values of the disorder range constant $h=5,10$, \corr{which exhibit plateaus,} the trend of \corr{$D$} is $1/r_c$ as happened in the case of the 3D plot.
\corr{Finally, in Fig.~\ref{fig:2_exc_2}(b) we show a 2D plot of the complete delocalization time $\tau$ as a function of the collision rate $r_c$ of Fig.~\ref{fig:2exc_3}, where the shape parameter $\nu$ and the disorder strength vary as indicated in the legend. In agreement with the 3D version, it is possible to recognize the two regions: the one without the plateaus (1) with small values of $\tau$; and the one with the formation of plateaus corresponding to the regimes (2) and (3), which has higher values of $\tau $. In addition, it is possible to see here that, for $r_c > 5$, as the system-environment interactions become more frequent, the dynamics is slowed down, leading to a gradual increase of the complete delocalization time $\tau$. This is a manifestation of the Zeno effect, also discussed in Sec.~\ref{One_Exc} and App.~\ref{App_Imb}. }

\begin{figure} [H]
    \centering
    \includegraphics[scale =0.23]{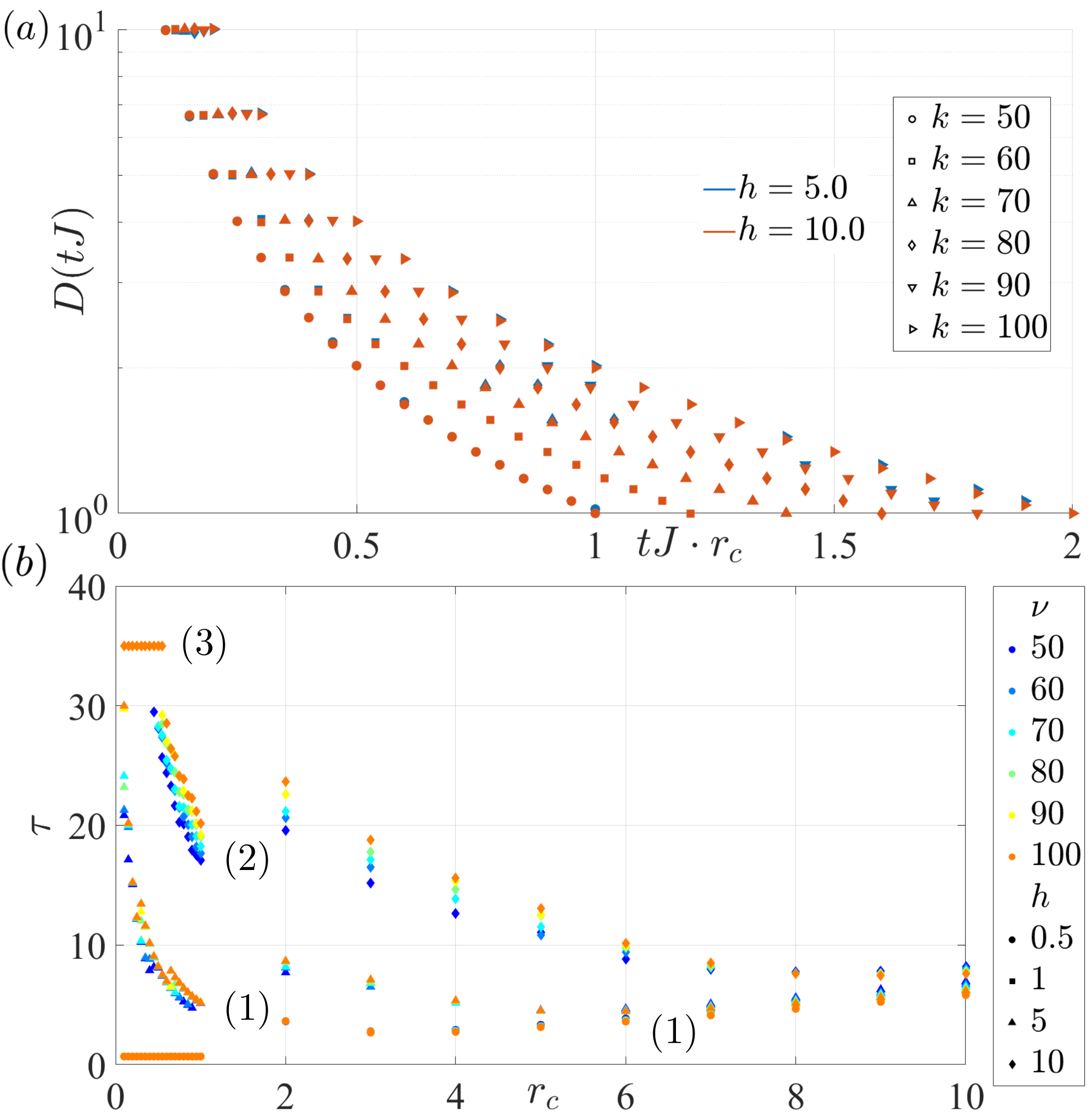}
    \caption{\corr{2D versions of the plots in Sec.~\ref{2_exc}.} \corr{(a) A semi-log version of the Duration} \corr{$D$} of the plateaus in terms of tJ vs the product $tJ \ r_c$, \corr{with  $h = 5, 10$ varying the shape parameter as in the legend.} \corr{(b) Complete delocalization time $\tau$ in Fig.~\ref{fig:2exc_3} vs the collision rate $r_c$. The shape parameter $\nu$ and the disorder strength $h$ vary with the color and shape legend, respectively. Also in this 2D version it is possible to recognize the two regions: without the plateaus with small values of $\tau$ (1); with the formation of plateaus corresponding to the regimes (2) and (3), which has higher values of $\tau $.} The simulations were performed for a spin chain of $N=20$ sites with two excitations initially separated by two spins\corr{, at fixed anisotropy $\Delta = 2.5$}, with $M = 250$ trajectories and timestep $dt = 0.02$. Final time of the simulations is $tJ = 30$}
    \label{fig:2dplot}
\end{figure}

\newpage
\bibliography{main}
\end{document}